\documentclass[aps,pra,twocolumn,superscriptaddress,longbibliography]{revtex4-1}%
\usepackage{graphicx}
\usepackage{float}
\usepackage{dcolumn}
\usepackage{bm,color}
\usepackage{amsmath,amssymb,dsfont,amstext,amsfonts}
\usepackage{extarrows}
\usepackage[colorlinks=true,linkcolor=blue,urlcolor=blue,citecolor=blue]{hyperref}
\usepackage{xcolor}
\usepackage{wasysym}
\usepackage{mathtools}
\usepackage{bbold}
\usepackage{tikz}
\usepackage{multirow}
\usepackage{subcaption}
\usepackage{caption}
\usepackage[normalem]{ulem} 
\usepackage{color}

\usepackage[colorlinks=true]{hyperref}  
\hypersetup{
    bookmarks=true,         
    unicode=false,          
    pdftoolbar=true,        
    pdfmenubar=true,        
    pdffitwindow=false,     
    pdfstartview={FitH},    
    pdfsubject={},   
    pdfcreator={},   
    pdfproducer={}, 
    pdfkeywords={} {} {}, 
    pdfnewwindow=true,      
    colorlinks=true,       
    linkcolor=blue, 
    citecolor=blue,        
    filecolor=magenta,      
    urlcolor=blue,           
	breaklinks=true
} 

\def\bs#1{\boldsymbol{#1}}

\newcommand{\sect}[1]{\vspace{0.3em}{\it #1.}---}

\makeatletter
\newcommand{\subalign}[1]{%
  \vcenter{%
    \Let@ \restore@math@cr \default@tag
    \baselineskip\fontdimen10 \scriptfont\tw@
    \advance\baselineskip\fontdimen12 \scriptfont\tw@
    \lineskip\thr@@\fontdimen8 \scriptfont\thr@@
    \lineskiplimit\lineskyip
    \ialign{\hfil$\m@th\scriptstyle##$&$\m@th\scriptstyle{}##$\hfil\crcr
      #1\crcr
    }%
  }%
}
\makeatother

\begin{document}
\title{Projected spin texture as a bulk indicator of fragile topology}
\author{Gunnar F. Lange}
\email{gfl25@cam.ac.uk}
\affiliation{TCM Group, Cavendish Laboratory, University of Cambridge, J.J.~Thomson Avenue, Cambridge CB3 0HE, United Kingdom}
\author{Adrien Bouhon}
\email{adrien.bouhon@gmail.com}
\affiliation{TCM Group, Cavendish Laboratory, University of Cambridge, J.J.~Thomson Avenue, Cambridge CB3 0HE, United Kingdom}
\affiliation{Nordic Institute for Theoretical Physics (NORDITA), Stockholm, Sweden}
\author{Robert-Jan Slager}
\email{rjs269@cam.ac.uk}
\affiliation{TCM Group, Cavendish Laboratory, University of Cambridge, J.J.~Thomson Avenue, Cambridge CB3 0HE, United Kingdom}

\date{\today}
\begin{abstract}
We study the relationship between projected momentum-space spin textures and Wilson loop winding, proving a map between band topology of and spin topology in certain restricted symmetry settings, relevant to fragile topology. Our results suggest that the spin gap may act as a smoking gun bulk indicator for fragile topology within specific scenarios. 
\end{abstract}

\maketitle
\sect{Introduction}
The concept of spin has played a perpetually significant role in physics. With the rise of topological materials this role was reinforced as paradigmatic time reversal (TRS) protected topological insulators \cite{Rmp1,Rmp2} thrive on having spinful TRS. Similarly, winding numbers associated with real space spin textures, such as Skyrmions \cite{Skyrmion}, have garnered significant attention over the years, as have Rashba/Dresselhaus spin-orbit coupling (SOC) terms in momentum space, which display (pseudo-)spin momentum locking~\cite{dresselhaus2007group}. These terms usually arise in a two-band $\boldsymbol{k}\cdot \boldsymbol{p}$ description in presence of strong SOC, describing a twofold degeneracy at a certain momentum. Generally, however, the topology of a gapped band structure is not captured by a purely local analysis, but instead requires comparing multiple momentum space points or paths (e.g. Wilson loops~\cite{bouhon2019wilson,Alex_BerryPhase, wi1}), leading to the consideration of spin textures across Brillouin zones (BZ). 

Whilst spin is a fundamental parameter in band topology, the specific role of spin textures are yet to be fully explored~\cite{Schrunk2022}. Such relations can however be motivated by specific examples. In a quantum spin Hall (QSH) phase preserving TRS for example, the relevant topological invariant quantifying the winding of the spin sector is the spin Chern number. This spin Chern number was shown to be robust as long as the spin gap does not close~\cite{Prodan_spin_chern} and in close correspondence with the Kane-Mele band-topological invariant~\cite{KaneMeleZ2}. In addition, spin plays an important role in the irreducible representations (IRREPs) that determine the gluing conditions~\cite{clas3}, which in turn fix the momentum space configurations considered in recent symmetry-based schemes such as topological quantum chemistry~\cite{Bradlyn2017} or symmetry indicators~\cite{clas4}.

Recently, unified relations between band topology and spin textures have received reinvigorated interest as reflected in work on describing spin-textures around higher degeneracies \cite{Barry_SML}, as well as moving away from local in momentum space (e.g. $\boldsymbol{k}\cdot \boldsymbol{p}$) descriptions, to a global description taking into account space-group symmetries \cite{Okuma2018} and spin textures at various points in the BZ \cite{Gatti2020,BZ_journey,Lin2022}. Similarly, we note that in Refs.~\cite{bouhon2019wilson, Bouhon2021} a relation between certain fragile topology and spin-locking processes has been previously reported. Given these appealing indications, we here discuss a symmetry setting where, rather than coexisting, there is a direct mapping between fragile topological phases and spin topology. In essence we are, inspired by the above-mentioned symmetry based classifications schemes, concerned with the question 
if there are connectivity rules amongst spin bands and whether they impact band topology.
Encouragingly, we are able to identify spin textures, and in particular the spin gap, as a bulk signature of fragile topological phases~\cite{Ft1,bouhonGeometric2020}. This is useful, as fragile phases are generically difficult to diagnose~\cite{Peri797}. 
Concretely we find:
\begin{itemize}
    \item In the presence of $C_{2}\mathcal{T}$ symmetry, with $[C_{2}\mathcal{T}]^2=1$, we can define a $\mathbb{Z}_2$ topological invariant, distinct from the spin Chern number, that relates to a superconducting symmetry class and characterizes the (vanishing of) the spin gap over the BZ.

    \item This invariant relates to the electronic band structure in two-band subspaces corresponding to a fragile split elementary band representation (EBR) \cite{Ft1,Zak_EBR4,bouhon2019wilson, Bradlyn2017} within specific symmetry settings, giving a bulk topological classification perspective of fragile electronic band topology. We find that whenever the spin invariant is trivial we necessarily have a fragile electronic phase.
   
   \item We discuss possible realistic experimental schemes to measure the spin gap and thus the $\mathbb{Z}_2$ invariant.
\end{itemize}

 This paper proceeds by introducing these ideas from a general point of view and subsequently applying and retrieving our findings in the context of two models: a hexagonal model having TRS (non-magnetic) and a tetragonal model with non-symmorphic TRS (magnetic).

\sect{(Pseudo-)spin connectivity and parallel transport}
We begin by defining spin textures in 2D systems with SOC that lack inversion symmetry. In such systems, spin is not a good quantum number across  the BZ, and we cannot straightforwardly associate a spin degree of freedom (DOF) to the bands. Even in the presence of SOC, however, there are $n$-fold degeneracies in the band structure to which we can associate an $\frac{n-1}{2}$ pseudo-spin DOF \cite{Barry_SML}. If such a degeneracy occurs at a high-symmetry point (HSP) $\boldsymbol{k}_0$, whose little-group contains a rotation axis perpendicular to the system's basal plane, then this gives a natural spin-up/spin-down quantization axis for the pseudo-spin. Namely, there exists a canonical gauge choice at the degeneracy, consisting of the pseudo-spin eigenstates  that diagonalize the rotation symmetry representation.
We will mostly focus on twofold degeneracies which are spanned by a pseudo-spin $\frac{1}{2}$ DOF. In this setting, the above gauge fixing reduces the gauge freedom at the degeneracy from $\mathrm{U}(2)$ to $\mathrm{U}(1)\times \mathrm{U}(1)$. The presence of (symmorphic or nonsymmorphic) time reversal symmetry (TRS), see below, is then used to further reduce the gauge freedom to $\mathrm{U}(1)$ by imposing that the pseudo-spins be the image of each-other under TRS, i.e. $\mathcal{T}(\chi_{\uparrow},\chi_{\downarrow}) \propto (\chi_{\uparrow},\chi_{\downarrow}) \sigma_{y} \mathcal{K}$, where $\mathcal{K}$ is the complex conjugation.

To explore the relationship between the (pseudo-)spin states at the degenerate points and topology, we need to consistently compare (pseudo-)spins at various points in the BZ in a gauge-controlled fashion. To attain gauge-consistent bands associated with (pseudo-)spin, $(|\chi_{\uparrow}, \boldsymbol{k}\rangle,| \chi_{\downarrow}, \boldsymbol{k}\rangle)$, we start from (pseudo-)spin eigenstates at a degenerate HSP $\boldsymbol{k}_0$ written in the above canonical gauge. At $\boldsymbol{k}_0$, we fix the remaining gauge phase factor of the eigenstates, and parallel transport these states from the degeneracy at $\boldsymbol{k}_0$ to some other $\boldsymbol{k}$ \footnote{Explicitly making such a construction across the {\it entire} BZ whilst preserving symmetries may be challenging, though it is possible for some symmetry classes as discussed in Ref.~\cite{Vanderbilt_smooth_gauge}. We will only be interested in the states along specific lines in the BZ, however, where such a construction can easily be performed.}. We will see that this construction is related to how the eigenvalues of the projected spin operator are used to disentangle degenerate bands in the QSH \cite{Prodan_spin_chern}.  We note that these parallel-transported states are not generically eigenstates of the Hamiltonian away from the degeneracy.

If $\boldsymbol{k}^*$ is another HSP whose point-group contains the same rotation axis as $\boldsymbol{k}_0$ then, if no gap closings with external bands occur, the parallel-transported states will again be eigenstates of the Hamiltonian at $\boldsymbol{k}^*$. We find numerically that the resulting states are also rotation eigenstates at $\boldsymbol{k}^*$, although a general proof of this property is still of active interest~\cite{bouhon2019wilson}. We conclude that the (pseudo-)spin eigenvalues, i.e. spin-up or spin-down, are good quantum numbers at both degeneracies. It is then meaningful to compare the (pseudo-)spin eigenvalue of the parallel-transported states at the HSPs $\boldsymbol{k}^*$ and $\boldsymbol{k}_0$ 
and consider whether it is preserved, i.e. indicating a parallel spin configuration, or reversed, i.e.  indicating a spin-flip configuration. This method of connecting (pseudo-)spin eigenvalues at distinct HSP across the Brillouin zone gives a characterization of the parallel-transported bands which is reminiscent of the above-mentioned symmetry-based classification methods~\cite{Clas1,Clas2,clas3,Bradlyn2017,clas4,mltop,mtqc,mSI} for Bloch bands. We stress however that the characterization in terms of the connectivity of pseudo-spin eigenvalues at degeneracies in the BZ is \textit{a priori} independent of these methods as we are concerned with how these (pseudo-)spin eigenvalues connect for a \textit{fixed} set of irreducible representations (IRREPs) at the degeneracies, and this spin-connectivity can generically change even when the IRREPs remain fixed. 

\sect{Spin bands}
In what follows we assume that we are dealing with twofold-degeneracies at which the spin-up and spin-down DOF are good quantum numbers. It turns out, moreover, that the Bloch eigenstates of the systems we consider realize the bare electron spin at these degeneracies.
To compare the spins, we could naively define the spin operator in the Bloch eigenframe $\{|u_n, \boldsymbol{k}\rangle\}_{n}$ (i.e. the Bloch energy eigenvectors) as \cite{Okuma2018}
\begin{equation}\label{eq:single_spin_band}
    \boldsymbol{\hat{S}}_s(\boldsymbol{k})_{n}= \langle u_n,\boldsymbol{k}|\hat{\boldsymbol{S}}| u_n,\boldsymbol{k}\rangle,
\end{equation}
where $\hat{\boldsymbol{S}}=\boldsymbol{n}\cdot \boldsymbol{\hat{\sigma}}$ is the spin operator along $\boldsymbol{n}$ in the basis of the Hamiltonian (e.g. tracing over orbital DOFs). This is the operator relevant for spin-ARPES measurements. However, this operator is not invariant under mixing of the Bloch eigenvectors of the occupied bands, as required by the parallel transport gauge. We therefore consider a multi-band generalization of Eq.~\eqref{eq:single_spin_band}, the projected spin operator, 
\begin{equation}\label{eq:projected_spin_def}
    \boldsymbol{\hat{S}}_P(\boldsymbol{k}) = P(\boldsymbol{k})\hat{\boldsymbol{S}}P(\boldsymbol{k}),
\end{equation}
considered in Refs.~\cite{Prodan_spin_chern,Lin2022}. Here $P(\boldsymbol{k})=\sum_{n=1}^{N_{\text{occ.}}}\vert u_n,\boldsymbol{k}\rangle \langle u_n,\boldsymbol{k} \vert = \mathcal{U}(\boldsymbol{k}) \mathcal{U}(\boldsymbol{k})^{\dagger}$ is the projector onto the occupied bands, where $\mathcal{U}(\boldsymbol{k}) = (|u_1,\boldsymbol{k}\rangle, |u_2,\boldsymbol{k}\rangle, \dots, |u_{N_{\mathrm{{occ}}}},\boldsymbol{k}\rangle)$ is the rectangular matrix of the occupied Bloch eigenvectors  (we assume $N_{\mathrm{occ}} = 2$ in what follows). Diagonalizing this operator allows us to define spin bands across the BZ. We note that the projected spin operator is gauge-invariant with respect to mixing of the occupied bands eigenvalues. At HSPs this operator has quantized eigenvalues $\pm \frac{1}{2}$, and two zero eigenvalues due to the two unoccupied bands (i.e. the projector $P(\bs{k})$ has rank 2, see the reduced spin operator below). When the two occupied spin bands are gapped, there exists a continuous frame with the same spin eigenvalue at $\Gamma$ and K. This is precisely the parallel transported frame constructed above. When the spin gap closes instead, it indicates an exchange of spin eigenvalues across the Brillouin zone. Furthermore, when the spin spectrum is gapped, there is a one-to-one correspondence between the winding of the Wilson loop of the occupied bands and the spin Chern number \cite{Prodan_spin_chern,Lin2022}, which we consider later. We conclude that the parallel transported frame and the spectral decomposition of the projected spin operator are complementary ways to relate the spin texture to electronic band topology.


\sect{Symmetry of the spin operator}
The existence of topological phases crucially depend on symmetries. We are here primarily interested in 2D systems with (symmorphic or non-symmorphic \cite{Bouhon2021}) time-reversal symmetry (TRS) $\mathcal{T}$ and out-of plane twofold rotation symmetry $C_{2z}$ with their product satisfying $[C_{2z}\mathcal{T}]^2 = +1$ (satisfied for both examples considered below), which we note has direct consequences for defining multi-gap invariants such as Euler class~\cite{bouhon2019wilson,bouhon2019nonabelian,AnEuler,BJY_nielsen,bouhonGeometric2020, unal_2020quench, Jiang2021}. The out-of plane rotation gives a natural spin quantization axis, and in what follows we focus mostly on the $\hat{S}_z$ operator. The operators $\mathcal{T}$ and $C_{2z}$ acts on the spin operator $\hat{\boldsymbol{S}}$ as
\begin{equation}\label{eq:T_spin}
\hat{\boldsymbol{S}}\xrightarrow[]{\mathcal{T}} -\hat{\boldsymbol{S}},
\end{equation}
\begin{equation}\label{eq:C2_spin}
\{\hat{S}_x,\hat{S}_y, \hat{S}_z\}\xrightarrow[]{C_{2z}}\{-\hat{S}_x, -\hat{S}_y, \hat{S}_z\}.
\end{equation}
Combining these transformations, and noting that $C_{2z}\mathcal{T}$ is local in momentum, it follows that the single spin operator $\hat{S}_{s,z}(\boldsymbol{k})_n=0$ away from degeneracy, suggesting that in this symmetry setting, spin-ARPES will be unable to detect the spin texture.

Let us note the symmetry action of $C_{2z}\mathcal{T}$ on the projector
\begin{equation}
    P(\boldsymbol{k})  \rightarrow  \hat{U}_{C_{2z}\mathcal{T}} P(\bs{k})^* \hat{U}_{C_{2z}\mathcal{T}}^{\dagger} ,
\end{equation}
where $\hat{U}_{C_{2z}\mathcal{T}}\mathcal{K}$ is the representation of $C_{2z}\mathcal{T}$ in the orbital basis, and the action of $C_{2z}\mathcal{T}$ symmetry on the Bloch eigenvectors 
\begin{equation}\label{eq:C2T_occ_main_text}
    \hat{U}_{C_{2z}\mathcal{T}} \mathcal{U}(\bs{k})^*  = \mathcal{U}(\bs{k})  \breve{W}_{C_{2z}\mathcal{T}}(\bs{k}) .   
\end{equation}
Working now with the projected spin-operator in the $z$-direction, Eq.~\eqref{eq:projected_spin_def}, we find the symmetry action (see Appendix~\ref{ap:C2T_spin}) 
\begin{equation}
    \hat{S}_{P,z}(\boldsymbol{k}) \xrightarrow []{C_{2z}\mathcal{T}} -\hat{U}_{C_{2z}\mathcal{T}}(\bs{k})\hat{S}_{P,z}(\boldsymbol{k})^*\hat{U}_{C_{2z}\mathcal{T}}(\bs{k})^{\dagger}.
\end{equation}

Because we require that $[C_{2z}\mathcal{T}]^2 = +1$, there exists a basis choice that transforms $\hat{U}_{C_{2z}\mathcal{T}}$ to the identity matrix (which can be found using an Autonne-Takagi decomposition, see Refs.~\cite{bouhon2019nonabelian, Peng2021} and Appendix~\ref{ap:C2T_spin}), such that $C_{2z}\mathcal{T}$ symmetry implies
\begin{equation}
    \hat{S}_{P,z}(\boldsymbol{k}) = -\hat{S}_{P,z}(\boldsymbol{k})^*,
\end{equation}
i.e. $\hat{S}_{P,z}$ is purely imaginary and off-diagonal in the new basis. Thus, in the presence of $C_{2z}\mathcal{T}$, the projected spin operator has particle-hole symmetry which is local in $\boldsymbol{k}$ and squares to $+1$. This places this operator in the superdconducting AZ+I (Altland-Zirnbauer + inversion) class D, which is characterized by the zeroth homotopy group $\pi_0=\mathbb{Z}_2$~\cite{tomas}. 

\sect{Topology of the spin operators}
To compute this nontrivial homotopy charge explicitly, we work in the real basis where the projected spin operator is purely off-diagonal. Then, the topological invariant is given by  \cite{tomas} $z_2^s = $sign(Pf[i$\hat{S}_{P,z}(\boldsymbol{k})])$, where Pf is the Pfaffian. This invariant can only change when there is a gap closing in the spin spectrum, which by $C_{2z}{\mathcal{T}}$ symmetry must happen when the eigenvalues of the projected spin operator cross $0$, thus implying an exchange of spin eigenvalues of the parallel transported frame from $\Gamma$ to K.

Note, however, as mentioned above, that the projected spin operator in systems with $N$ bands generically has $N-N_{\mathrm{occ}}$ zero modes, associated with the unoccupied bands \cite{Lin2022}. These zero modes force $z_2^s$ to be trivial. We can therefore instead work with the reduced spin operator,
\begin{equation}\label{eq:reduced_spin_def}
    \boldsymbol{S}^{r}_P(\boldsymbol{k})_{nm} = \langle u_n, \boldsymbol{k}|\hat{\boldsymbol{S}}|u_m, \boldsymbol{k}\rangle = [\mathcal{U}(\boldsymbol{k})^{\dagger}\hat{\boldsymbol{S}}_P \mathcal{U}(\boldsymbol{k})]_{nm},
\end{equation}


As discussed in \cite{Lin2022}, the spectrum of the reduced spin operator is identical to that of the projected spin operator, except that it excludes the zero modes associated with the unoccupied bands. The reduced spin operator is gauge covariant (see Appendix~\ref{ap:C2T_spin}), so that the eigenvalues are gauge invariant.  We repeat the above symmetry analysis for the $z$-component of the reduced spin operator, and find (see the derivation in Appendix~\ref{ap:C2T_spin})
\begin{equation}\label{eq:C2T_red_spin_main}
    \hat{S}^r_{P,z}(\bs{k}) = -\breve{W}_{C_{2z}\mathcal{T}}(\bs{k}) \hat{S}^r_{P,z}(\bs{k})^* \breve{W}_{C_{2z}\mathcal{T}}(\bs{k})^{\dagger} ,
\end{equation}
where we have used the representation of $C_{2z}\mathcal{T}$ in the occupied Bloch eigenvectors, Eq.~\eqref{eq:C2T_occ_main_text}. As we discuss in Appendix~\ref{ap:C2T_spin}, upon a change to the real basis $\breve{W}_{C_{2z}\mathcal{T}}(\bs{k})$ goes to the identity, so that Eq.~\eqref{eq:C2T_red_spin_main} again makes the $z$-component of the projected spin operator imaginary and off-diagonal. This allows us to compute the above topological invariant from the Pfaffian of the reduced spin operator in the real basis. 
By computing in the parallel-transport gauge defined above, we can therefore define a (generically) non-trivial $\mathbb{Z}_2$ invariant associated with the spin bands,
\begin{equation}\label{eq:reduced_pfaffian_def}
    z_{2,\mathrm{red}}^s = \mathrm{sign}(\mathrm{Pf}[i\hat{S}^r_{P,z}(\boldsymbol{k})]).
\end{equation}
Crucially, this invariant {\it can only change when the spin gap closes}, which should happen along a spin nodal ring surrounding one of the HSPs. In the case of a two-band parallel-transported frame discussed above, the nontrivial spin invariant indicates that the spin eigenvalues cross zero an odd number of times between the twofold degeneracies of the different HSPs. This invariant therefore captures the spin connectivity between degeneracies discussed above. We next show that this spin connectivity is in direct correspondence with the fragile topology of the electronic bands.

\sect{Pseudo-spin and band topology}
A standard way to diagnose band topology is by using the Wilson loop operator $W(\boldsymbol{k})$~\cite{bouhon2019wilson,Alex_BerryPhase, wi1}. 
We have previously established an analytic relationship between the Wilson loop winding and the pseudo-spin structure in two specific symmetry settings. Namely, in Ref.~\cite{bouhon2019wilson} we studied the four-band EBR of the honeycomb lattice populated with a $p_z$ orbital and two $1/2$-spins on each sub-lattice site, assuming the hexagonal space group P6mm$1'$, i.e. with the point group $C_{6v}$ and TRS. We showed that when the EBR is split into two occupied and two unoccupied bands, the two-band subspace that hosts twofold degeneracies at $\Gamma$ and $K$ exhibits a rigid relationship between the winding of Wilson loop $W$ and its spin structure. More precisely, we found that all phases with SOC that are adiabatically connected to gapped phases with inversion symmetry (i.e. with zero Rashba SOC) only admit spin textures with aligned spins at $\Gamma$ and K and exhibit a Wilson loop winding $W\in\pm 1+3\mathbb{Z}$. Moreover, we found phases with a spin-flip configuration, from $\Gamma$ to K, that are not adiabatically connected to inversion-symmetric phases and only admit a Wilson loop winding $W\in 3\mathbb{Z}$ (in practice, we only found models of such phases with $W=0$). We recall that an odd Wilson loop winding in this system implies a nontrivial $\mathbb{Z}_2$ QSH phase, while all other phases are fragile topological. 
In addition, in Ref.~\cite{Bouhon2021}, we found similar results in a tetragonal, antiferromagnetic space group P$_C$4, i.e. with point group $C_4$ and non-symmorphic TRS $\mathcal{T}\{E\vert [\bs{t}_1+\bs{t}_2]/2\}$. Specifically, we find the relationships shown in Tab.~\ref{tab:pseudo_spin_Wilson}.
\begin{table}[ht!]
    \centering
    \begin{tabular}{c c|c | c}
          \multirow{2}{*}{$\Gamma$} & \multirow{2}{*}{K/M} &   \multicolumn{2}{c}{Wilson loop winding $W$}\\
          & & $C_{6v}$+$\mathcal{T}$ & $C_4$+$\mathcal{T}\{E\vert [\bs{t}_1+\bs{t}_2]/2\}$ \\
         \hline
          $\{\uparrow, \downarrow\}$  & $\{\uparrow, \downarrow\}$ &$\pm 1 + 3\mathbb{Z}$ & $1+2\mathbb{Z}$\\
          $\{\uparrow, \downarrow\}$  & $\{\downarrow, \uparrow\}$ &$3\mathbb{Z}$ & $2\mathbb{Z}$\\
    \end{tabular}
    \caption{Relationship between the spin eigenvalues ($\uparrow$ for $+1/2$, $\downarrow$ for $-1/2$) at $\Gamma$ and K/M of the parallel-transported pseudo-spin eigenframe and Wilson loop winding for a hexagonal/tetragonal system in the specific symmetry setting described in Refs.~\cite{bouhon2019wilson,Bouhon2021}.}
    \label{tab:pseudo_spin_Wilson}
\end{table}

Crucially, we find in this work that the two-band subspaces with zero Wilson loop winding, i.e. \textit{necessarily} hosting a spin-flip configuration between $\Gamma$ and K/M, precisely exhibit a change in the $z_{2,\mathrm{red}}^s$ invariant, Eq.~\eqref{eq:reduced_pfaffian_def}, so that we obtain in the hexagonal case:
\begin{equation}
    \begin{aligned}
        z_{2,\mathrm{red}}^s = 0\  \mathrm{mod\  }2 &\iff W \in \pm  1 +3\mathbb{Z},\\
        z_{2,\mathrm{red}}^s = 1\  \mathrm{mod\  }2 &\iff W = 0 + 3\mathbb{Z}.
    \end{aligned}
\end{equation}
A similar result can be verified for the tetragonal case. We therefore see that the $\mathbb{Z}_2$ spin invariant acts as a complement to the electronic topology: {\it the two-band subspace of a split EBR with trivial Wilson-loop winding in this specific symmetry setting is only allowed when there is a non-trivial $\mathbb{Z}_2$ spin invariant, implying a stable spin-spectrum nodal ring. Conversely, a trivial $\mathbb{Z}_2$ spin invariant implies non-zero Wilson loop winding}. This constitutes a main finding of this paper and we illustrate these results for two phases with different Wilson-loop windings in the hexagonal case in Fig.~\ref{fig:C6_combined}. Similar results for the antiferromagnetic tetragonal counterpart are detailed in Appendix~\ref{ap:C4_results}.
\begin{figure*}[ht!]
    \centering
    \includegraphics[width=\textwidth]{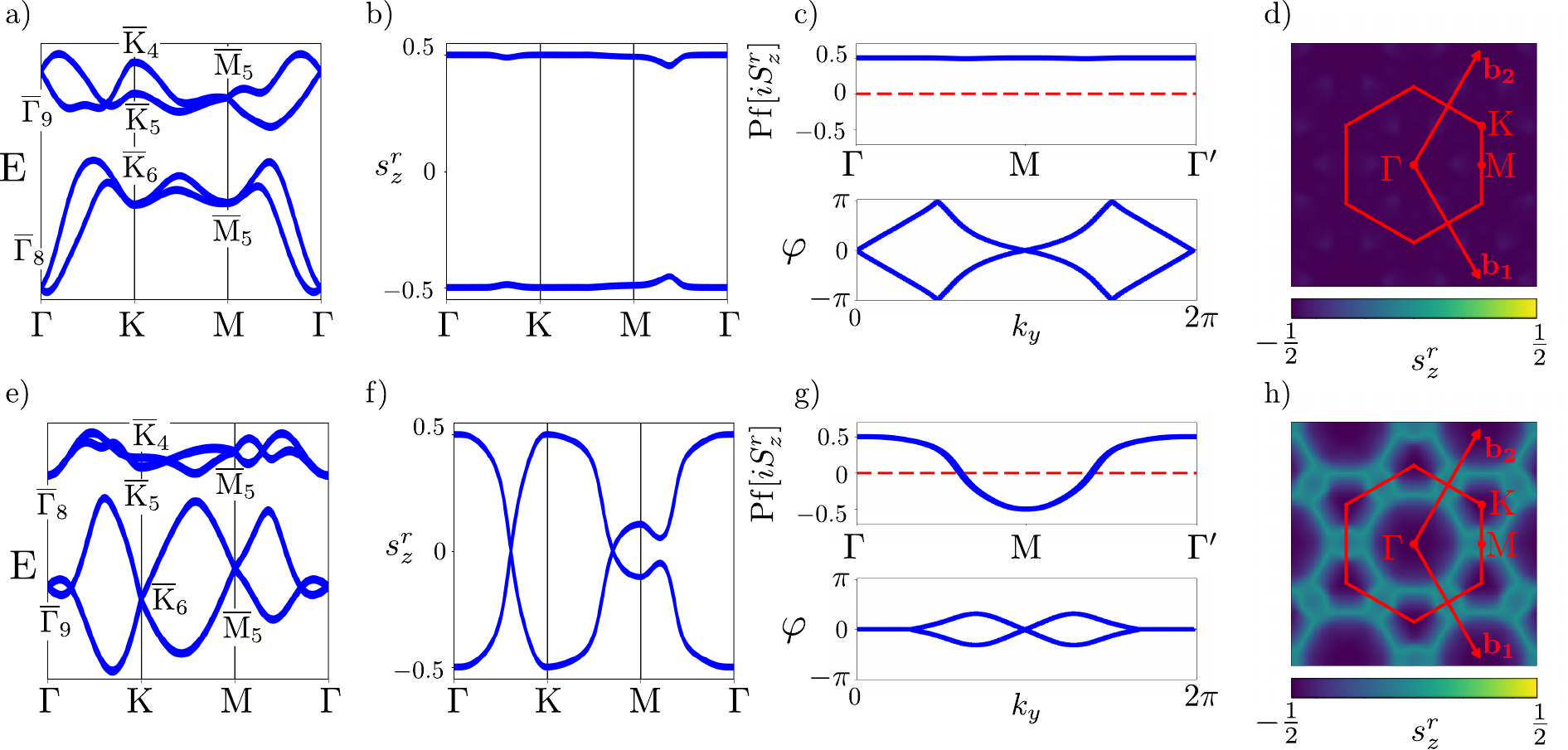}
    \caption{Relevant figures for the P6mm$1$' symmetric case. In a)-d) we show the fragile case (where the Wilson loop of the occupied and unoccupied bands winds), whereas e)-h) shows the trivial case (where the Wilson loop of the occupied bands does not wind). In a) and e) we show the band structures, with IRREPs of P6mm$1$' indicated. In b) and f) we show the eigenvalues of the reduced spin operator along high-symmetry paths of the BZ. Note that spin is not a good quantum number at M, but is a good quantum number at $\Gamma$ and K. In c) and g), we show the Pfaffian and Wilson loop of the occupied bands over the BZ, with $\Gamma' = \Gamma+\boldsymbol{b}_1+\boldsymbol{b}_2$.  Comparing the plots, we see that a non-trivial Wilson loop corresponds to a trivial Pfaffian invariant and vice-versa. Finally in d) and h) we show the spin expectation value over the BZ of the lowest spin band [as shown along high-symmetry lines in b) and f)]. This shows that the case with trivial Wilson loop has spin nodal lines. For more details on the models, see Appendix~\ref{ap:model_details} and Ref.~\cite{bouhon2019wilson}.}
    \label{fig:C6_combined}
\end{figure*}

\sect{Experimental detection of spin operator}
As discussed above, spin-ARPES is generically not capable of resolving the projected spin texture. However, if the spin invariant $z_{2,\mathrm{red}}^s$ is non-trivial, then the spin gap necessarily vanishes somewhere along the relevant high-symmetry line. Conversely, this implies that whenever the reduced spin operator is gapped across the entire BZ, we necessarily have a non-trivial winding of the Wilson loop within the split EBR context and symmetry settings we are considering. In general, the spin gap can close at points away from the high-symmetry line, or it can close twice along the line, without changing the spin invariant and therefore the Wilson loop quantization. Thus, when the spin is gapless, we \textit{may} still be in a phase where the Wilson loop winds (note, however, that Ref.~\cite{Lin2022} showed that the spin gap is perturbatively stable). However, when the spin spectrum is gapped, we are \textit{necessarily} in a phase where the Wilson loop winds. 

We can therefore detect topology by measuring the gap in the projected spin spectrum. As pointed out in \cite{Lin2022}, the spin gap provides a bound on the excitation spectrum of a system when quenching with a $\hat{z}$-directed Zeeman field. Similarly, we also expect this gap to show up in the spin structure factor of neutron scattering experiments. For the Zeeman quenched case, the density of excited states is (as shown in Ref.~\cite{Lin2022}):
\begin{equation}
    \Gamma_{ex}= \sum_{\boldsymbol{k}\in \mathrm{BZ}}\mathrm{Tr}\big[P(\boldsymbol{k})\hat{S}_z[1-P(\boldsymbol{k})]\hat{S}_zP(\boldsymbol{k})\big].
\end{equation}
Importantly, we find that $\Gamma_{ex}$ is an order of magnitude larger in the case where the spin spectrum is gapless.

\sect{Relation to spin Chern number}
When the gap in the projected spin operator is open across the BZ, one can compute the Wilson loop of the spin bands. This corresponds to the spin-Chern number~\cite{Prodan_spin_chern}. We define the Chern number of the upper/lower spin bands as $C^{\pm}$, and define the spin-Chern number as $C_s = C^+-C^-$. As discussed in Appendix \ref{ap:Spin_Wilson_loops} and shown in Fig.~\ref{fig:spin_wilson_loop_winding}, we find in the non-magnetic hexagonal case that $C_s = 4$, whereas in the magnetic tetragonal case, $C_s = 2$. In non-magnetic systems, the spin Chern number is related to the Kane-Mele $\mathbb{Z}_2$ invariant as $\nu_{\mathrm{KM}} = (1/2) (C_s\ \mathrm{mod}\ 4)$.  Thus, the hexagonal case does not have a strong invariant associated with the electronic band structure for even winding of the Wilson loop, in agreement with "fragile" topology (in fact, Ref.~\cite{bouhon2019wilson} showed that the fragile topology with $W\in \pm 4+12 \mathbb{Z}$ can be trivialized by coupling to trivial bands, see also Fig.~\ref{fig:extended_WL}). In the magnetic case, this corresponds to having a QSH phase without TRS, which has also been investigated in Ref.~\cite{Yang2011,Xiao2022}, where it was found that the QSH edge states can be generically gapped when TRS is broken. We showed in Ref.~\cite{Lange2021} that the surface spectrum in the magnetic case agrees with that expected of a fragile topological insulator, and the IRREP content is also consistent with fragile topology, justifying the label of fragile topological phase.

\sect{Conclusions}
We have discussed a general scheme to investigate refined band topologies beyond symmetry indicators, and how it relates to (pseudo-)spin DOFs. We have further shown that, for certain symmetry settings, the existence of a gap in a certain spin operator directly indicates the presence of fragile topology. This provides a bulk indicator for fragile topology, and we discuss concrete experimental schemes to measure this spin gap.

\sect{Outlook}
An interesting remaining question concerns the more general relationship between the spin-invariant and the band topology, beyond two-band degeneracies. Note, in particular, that the $\mathbb{Z}_2$ invariant $z^s_{2,\mathrm{red}}$ is defined for any even number of bands in the occupied space, in any system with $C_{2z}\mathcal{T}$ symmetry. However, the pseudo-spin connectivities, and their relationship to Wilson loop were specifically derived in a two-band subspace. The connection with higher degeneracies, or additional subspaces, comprise an interesting future pursuit. Furthermore, it is interesting to note that we uncovered a similar relationship between Wilson loop windings and pseudo-spin orderings in two very different space groups. This points to a more generic structure. As Wilson loops are difficult to measure, such a relationships to an observable would be of particular interest.  We also note that our results can apply to any physical operator that can be associated with the degeneracies in the BZ.
Finally, it would also be interesting to more systematically consider the role that SOC, and the vanishing of spin-flip terms at certain HSPs, plays as well as the role of the spin Chern number in these systems and generalizations to different quantities such as sub-lattice winding numbers.

\begin{acknowledgements}
G.F.L is funded by the Aker Scholarship. We thank the Marie Sk\l{}odowska-Curie programme under EC Grant No. 84290, the Winton Programme for the Physics of Sustainability, a New Investigator Award, EPSRC grant
EP/W00187X/1 and Trinity College at the University of Cambridge 
\end{acknowledgements}

\newpage
\bibliography{CambridgeSpinTexture}
\appendix
\section{Symmetry of spin operators under $C_{2z}\mathcal{T}$}\label{ap:C2T_spin}

\subsection{Defining the spin operators}
We define the Bloch orbital basis in which the Bloch Hamiltonian is defined
\begin{equation}
    \vert \varphi_{\alpha},\bs{k} \rangle 
     = \dfrac{1}{\sqrt{N}} \sum\limits_{\bs{k}} e^{i \bs{k}\cdot (\bs{R}_m + \bs{r}_{\alpha})} \vert w_{\alpha},\bs{R}_m+\bs{r}_{\alpha} \rangle\,,
\end{equation}
with the atomic Wannier function $\langle \bs{r}\vert w_{\alpha},\bs{R}_m+\bs{r}_{\alpha} \rangle = w_{\alpha}(\bs{r}-\bs{R}_m-\bs{r}_{\alpha})$ for the $\alpha$-th atomic orbital. The Bloch eigenstates are then defined through
\begin{equation}
    \vert \psi_n,\bs{k} \rangle = 
    \vert \bs{\varphi},\bs{k} \rangle  [\mathcal{U}_t(\bs{k})]_n  \,,
\end{equation}
with $\mathcal{U}_t(\bs{k})$ the matrix of column Bloch eigenvectors $[\mathcal{U}_t(\bs{k})]_n=\vert u_n,\bs{k}\rangle$, i.e. 
\begin{equation}
    \mathcal{U}_t(\boldsymbol{k}) = (|u_1,\boldsymbol{k}\rangle ,|u_2, \boldsymbol{k}\rangle, \dots, |u_{N},\boldsymbol{k}\rangle).
\end{equation}

Where $N$ is the total number of bands, which we assume to be even. From Eq.~\eqref{eq:projected_spin_def}, the projected spin operator in the occupied Bloch band basis is given by:
\begin{equation}
\hat{\boldsymbol{S}}_P(\boldsymbol{k}) = P(\boldsymbol{k})\hat{\boldsymbol{S}}P(\boldsymbol{k}) = \sum_{n,m \in \mathrm{occ}} \langle u_n, \boldsymbol{k}|\hat{\boldsymbol{S}}|u_m \boldsymbol{k}\rangle |u_n, \boldsymbol{k}\rangle \langle u_m , \boldsymbol{k}|
\end{equation}
Where $\hat{\boldsymbol{S}}$ is the spin operator when tracing over all orbital DOFs and $|u_n, \boldsymbol{k}\rangle$ is the cell-periodic part of occupied band $n$. We can write a multiband gauge-transformation of the occupied bands as:
\begin{equation}
    |\tilde{u}_n, \boldsymbol{k}\rangle = \mathcal{U}(\bs{k})G(\bs{k})  
    = \sum_{m\in \mathrm{occ}} G_{mn}(\bs{k})|u_m, \boldsymbol{k}\rangle 
\end{equation}
with $G(\bs{k})\in \mathsf{U}(N_{\mathrm{occ}})$, and $\mathcal{U}(\boldsymbol{k})$ is the rectangular matrix of occupied Bloch eigenvectors
\begin{equation}
    \mathcal{U}(\boldsymbol{k}) = (|u_1,\boldsymbol{k}\rangle ,|u_2, \boldsymbol{k}\rangle, \dots, |u_{N_{\mathrm{occ}}},\boldsymbol{k}\rangle),
\end{equation}
which satisfies
\begin{equation}
    \begin{aligned}
        \mathcal{U}\mathcal{U}^{\dagger} &= P\\
        \mathcal{U}^{\dagger}\mathcal{U}&= \mathbb{1}_{N_{\mathrm{occ}}}\times  \mathbb{1}_{N_{\mathrm{occ}}}.
    \end{aligned}
\end{equation}
We find that the projection operator is gauge-invariant, so that the projected spin operator 
$\hat{\boldsymbol{S}}_P$ is also gauge-invariant. We are also interested in the reduced spin operator:
\begin{equation}
    \hat{\boldsymbol{S}}_P^r(\boldsymbol{k})_{nm} = \langle u_n, \boldsymbol{k}|\hat{\boldsymbol{S}}|u_m, \boldsymbol{k}\rangle = (\mathcal{U}^{\dagger}\hat{\boldsymbol{S}}_P\mathcal{U})_{nm}
\end{equation}
This operator is gauge-covariant, as under a gauge-transformation $G(\bs{k})\in \mathsf{U}(N_{\mathrm{occ}})$:
\begin{equation}
    \hat{\boldsymbol{S}}_P^r(\boldsymbol{k}) \rightarrow \tilde{\hat{\boldsymbol{S}}}_P^r(\boldsymbol{k}) =  G(\bs{k})^{\dagger} \boldsymbol{\hat{S}}_P^r(\boldsymbol{k}) G(\bs{k})
\end{equation}
So that the spin eigenvalues are gauge-invariant. We now wish to understand  how these operators behave under $C_{2z}\mathcal{T}$ symmetry.

\subsection{Action of $C_{2z}\mathcal{T}$}
Letting $\hat{U}_{\mathcal{T}}, \hat{U}_{C_{2z}}$ be the unitary actions of time-reversal and $C_{2z}$ respectively, we  define $\hat{U}_{C_{2z}\mathcal{T}} = \hat{U}_{C_{2z}}\hat{U}_{\mathcal{T}}$. We require that $(C_{2z}\mathcal{T})^2 = +1$, which gives that $\hat{U}_{C_{2z}\mathcal{T}}U_{C_{2z}\mathcal{T}}^{\dagger}=\mathbb{1}$ and $\hat{U}_{C_{2z}\mathcal{T}}\hat{U}_{C_{2z}\mathcal{T}}^{*}=\mathbb{1}$, so that $\hat{U}_{C_{2z}\mathcal{T}} = [\hat{U}_{C_{2z}\mathcal{T}}]^T$. We now derive the symmetry properties of the projected and reduced spin operator. We first give the action of the $C_{2z}\mathcal{T}$ symmetry on the occupied Bloch eigenstate basis, 
\begin{equation}
\begin{aligned}
    ^{C_{2z}\mathcal{T}}\vert \psi_n,\bs{k} \rangle &= ^{C_{2z}\mathcal{T}}\vert \bs{\varphi},\bs{k}\rangle [\mathcal{U}(\bs{k})]_n \,,\\
    &=  \vert \psi_m ,\bs{k}\rangle [\mathcal{U}(\bs{k})^{\dagger}]_m \hat{U}_{C_{2z}\mathcal{T}} [\mathcal{U}(\bs{k})^*]_n \mathcal{K}   \,,\\
    &= \vert \psi_m ,\bs{k}\rangle [\breve{W}_{C_{2z}\mathcal{T}}(\bs{k})]_{mn}  \mathcal{K} \,,
\end{aligned}
\end{equation}
where $\breve{W}_{C_{2z}\mathcal{T}}(\bs{k})$ is a $N_{\mathrm{occ}}\times N_{\mathrm{occ}}$ unitary matrix that represents the $C_{2z}\mathcal{T}$ symmetry on the occupied-band eigen-subspace under consideration (we assume that we can isolate the occupied subspace from the other bands). For later reference, it is also useful to rewrite it as an action on the occupied Bloch eigenframe
\begin{equation}
\label{eq_C2T_rep}
    \hat{U}_{C_{2z}\mathcal{T}} \mathcal{U}(\bs{k})^* = \mathcal{U}(\bs{k}) \breve{W}_{C_{2z}\mathcal{T}}(\bs{k})  \,.
\end{equation}

The action of $C_{2z}\mathcal{T}$ on the two-band projector is then
\begin{equation}
    P(\bs{k}) \rightarrow \hat{U}_{C_{2z}\mathcal{T}} P(\bs{k})^* \hat{U}_{C_{2z}\mathcal{T}}^{\dagger},
\end{equation}
Now note that:
\begin{equation}
\begin{split}
\hat{U}_{C_{2z}\mathcal{T}}^{\dagger} \hat{S}_x \hat{U}_{C_{2z}\mathcal{T}} = \hat{S}_x\\
\hat{U}_{C_{2z}\mathcal{T}}^{\dagger} \hat{S}_y \hat{U}_{C_{2z}\mathcal{T}} = \hat{S}_y\\
\hat{U}_{C_{2z}\mathcal{T}}^{\dagger} \hat{S}_z \hat{U}_{C_{2z}\mathcal{T}} = -\hat{S}_z
\end{split}
\end{equation}
E.g., in the orbital basis $C_{2z}\mathcal{T}$ flips the $z$-component of spin whilst leaving the others unchanged. Thus, the action on the projected spin-z operator is:
\begin{equation}\label{eq:ap_projected_spin_op}
\begin{aligned}
     \hat{S}_{P,z}(\bs{k}) &\rightarrow \hat{U}_{C_{2z}\mathcal{T}} P(\bs{k})^* (\hat{U}_{C_{2z}\mathcal{T}}^{\dagger}\hat{S}_z\hat{U}_{C_{2z}\mathcal{T}}) P(\bs{k})^* \hat{U}_{C_{2z}\mathcal{T}}^{\dagger} \\
    &= -\hat{U}_{C_{2z}\mathcal{T}} P(\bs{k})^* \hat{S}_z P(\bs{k})^* \hat{U}_{C_{2z}\mathcal{T}}^{\dagger}\\
    &= - \hat{U}_{C_{2z}\mathcal{T}} \hat{S}_{P,z}^* \hat{U}_{C_{2z}\mathcal{T}}^{\dagger} \,.
\end{aligned}
\end{equation}
The $C_{2z}\mathcal{T}$ symmetry then requires
\begin{equation}
\label{eq_C2T_constraint_P}
    \hat{S}_{P,z}(\bs{k}) = - \hat{U}_{C_{2z}\mathcal{T}} \hat{S}_{P,z}^*  \hat{U}_{C_{2z}\mathcal{T}}^{\dagger}\,.
\end{equation}
This implies that the z-component of the projected spin operator has a local in $\boldsymbol{k}$, antiunitary, antisymmetry, squaring to $+1$. Following \cite{tomas} this puts this operator in AZ+$\mathcal{I}$ symmetry class D, characterized by $\pi_0 = \mathbb{Z}_2$ and spectral symmetry around $s_z = 0$. 

To find an explicit expression for this $\mathbb{Z}_2$ topological invariant associated with the spin spectrum, we note that, due the properties of $\hat{U}_{C_{2z}\mathcal{T}}$ discussed previously, there exists a basis transformation (given by an Autonne-Takagi decomposition \cite{bouhon2019nonabelian}), such that $\hat{U}_{C_{2z}\mathcal{T}}=\mathbb{1}$. In this basis, $\hat{S}_{P,z}(\boldsymbol{k})$ is a purely imaginary, skew-antisymmetric matrix at all $\boldsymbol{k}$, from which it follows that Pf($i\hat{S}_{P,z}$) is well-defined. Then, in a smooth gauge, the Pfaffian evolves continuously, and only switches sign through a spin gap closing. It follows that $\mathrm{sign}[\mathrm{Pf}(i\hat{S}_{P,z})]$, evaluated at two different points, gives the required invariant. Note, however, that Pf($A$)$^2 = $Det($A$) and Det($\hat{S}_{P,z}) = 0$ because $\hat{S}_{P,z}$ always has  $N-N_{\mathrm{occ}}$ zero eigenvalues, associated with the eigenstates of $Q = 1-P$ (physically: the conduction bands). Thus, the $\mathbb{Z}_2$ invariant is always ill-defined for $\hat{S}_{P,z}$. 

To circumvent this issue, we instead look at the reduced spin operator, Eq.~\eqref{eq:reduced_spin_def}, where the zero eigenvalues are projected out. Rewriting the projected spin operator as
\begin{equation}
\label{eq_reduced_spin_op}
    \hat{S}_{P,z}(\bs{k}) = \mathcal{U}(\bs{k}) \hat{S}^r_{P,z}(\bs{k}) \mathcal{U}(\bs{k})^{\dagger}\,,
\end{equation}
we find from Eq.\,(\ref{eq_C2T_constraint_P}) and Eq.\,(\ref{eq_C2T_rep}) 
\begin{equation}
\begin{aligned}
     \hat{S}_{P,z}(\bs{k}) &\rightarrow  -\hat{U}_{C_{2z}\mathcal{T}} P(\bs{k})^* \hat{S}_z P(\bs{k})^* \hat{U}_{C_{2z}\mathcal{T}}^{\dagger}\\
    &= \hat{U}_{C_{2z}\mathcal{T}} \mathcal{U}(\bs{k})^* [-\mathcal{U}(\bs{k})^{\dagger}\hat{S}_z\mathcal{U}(\bs{k})]^* P(\bs{k})^* \\
    & = -\mathcal{U}(\bs{k}) \breve{W}_{C_{2z}\mathcal{T}}(\bs{k}) \hat{S}^r_{P,z}(\bs{k})^* \breve{W}_{C_{2z}\mathcal{T}}(\bs{k})^{\dagger} \mathcal{U}(\bs{k})^{\dagger}.
\end{aligned}
\end{equation}
Since $C_{2z}\mathcal{T}$ symmetry requires that the last line must equal $\hat{S}_{P,z}(\bs{k}) $ in the form of Eq.\,(\ref{eq_reduced_spin_op}), we finally get
\begin{equation}
    \hat{S}^r_{P,z}(\bs{k}) = -\breve{W}_{C_{2z}\mathcal{T}}(\bs{k}) \hat{S}^r_{P,z}(\bs{k})^* \breve{W}_{C_{2z}\mathcal{T}}(\bs{k})^{\dagger} .
\end{equation}
Which is very similar to the particle-hole symmetry in Eq.~\eqref{eq_C2T_constraint_P}, except that the symmetry operator now depends on $\boldsymbol{k}$, as a consequence of projecting out the zero eigenvalues. Note that:
\begin{equation}
   \breve{W}_{C_{2z}\mathcal{T}}(\bs{k})[\breve{W}_{C_{2z}\mathcal{T}}(\bs{k})]^* = \breve{W}_{C_{2z}\mathcal{T}}(\bs{k})[\breve{W}_{C_{2z}\mathcal{T}}(\bs{k})]^{\dagger} =\mathbb{1}\\
\end{equation}
Because $\hat{U}_{C_{2z}\mathcal{T}} = [\hat{U}_{C_{2z}\mathcal{T}}]^T$. 
It is therefore possible to perform an Autonne-Takagi basis change at each $\boldsymbol{k}$ to make $\hat{S}_{P,z}^r(\boldsymbol{k})$ antisymmetric and off-diagonal. This decomposition will however amount to a \textit{local} rotation of the basis, rather than a global rotation. This may induce additional topological structure. We now show that this does not happen when working in the basis that makes the Hamiltonian real.

To show this, we now perform the Autonne-Takagi transformation \cite{bouhon2019nonabelian}. Because $\hat{U}_{C_{2z}\mathcal{T}} = [\hat{U}_{C_{2z}\mathcal{T}}]^T$, we can write $\hat{U}_{C_{2z}\mathcal{T}} = V_{C_{2z}\mathcal{T}}D[V_{C_{2z}\mathcal{T}}]^{T}$ for a unitary $V_{C_{2z}\mathcal{T}}$ and a diagonal $D$. Defining $V_{\mathbb{R}} = \sqrt{D^*}V_{C_{2z}\mathcal{T}}^{\dagger}$, the Hamiltonian is made real by the basis change:
\begin{equation}
    H_{\mathbb{R}}(\boldsymbol{k}) = V_{\mathbb{R}} H(\boldsymbol{k})V_{\mathbb{R}}^{\dagger}
\end{equation}
The eigenstates of the Hamiltonian in this basis can be chosen to be real, as we numerically verified, and are related to the eigenstates in the original basis by: $|u_n^{\mathbb{R}}, \boldsymbol{k}\rangle = e^{i\varphi} V_{\mathbb{R}} |u_n, \boldsymbol{k}\rangle$, for some arbitrary gauge phase $\varphi$. Under $C_{2z}\mathcal{T}$:
\begin{equation}
\begin{split}
    H(\boldsymbol{k}) = V_{\mathbb{R}}^{\dagger}H_{\mathbb{R}}(\boldsymbol{k})V_{\mathbb{R}}  = \hat{U}_{C_{2z}\mathcal{T}}H^*(\boldsymbol{k})\hat{U}_{C_{2z}\mathcal{T}}^{\dagger}\\
    = \hat{U}_{C_{2z}\mathcal{T}}V_{\mathbb{R}}^TH_{\mathbb{R}}(\boldsymbol{k})V_{\mathbb{R}}^* \hat{U}_{C_{2z}\mathcal{T}}^{\dagger}
\end{split}
\end{equation}
And thus:
\begin{equation}
     H_{\mathbb{R}}(\boldsymbol{k}) = V_{\mathbb{R}} \hat{U}_{C_{2z}\mathcal{T}}V_{\mathbb{R}}^TH_{\mathbb{R}}(\boldsymbol{k})V_{\mathbb{R}}^* \hat{U}_{C_{2z}\mathcal{T}}^{\dagger} V_{\mathbb{R}}^{\dagger}
\end{equation}
Defining $\hat{U}_{C_{2z}\mathcal{T}, \mathbb{R}} = V_{\mathbb{R}}\hat{U}_{C_{2z}\mathcal{T}}V_{\mathbb{R}}^T$, we note that $\hat{U}_{C_{2z}\mathcal{T}, \mathbb{R}} = \sqrt{D^*}D\sqrt{D^*} = \mathbb{1}$, so that $C_{2z\mathcal{T}}$ acts diagonally in the real basis. We can similarly define the projected spin operator in occupied bands of the real basis:
\begin{equation}
\hat{\boldsymbol{S}}_{P,\mathbb{R}} = \sum_{n,m\in \mathrm{occ}} \langle u^{\mathbb{R}}_n, \boldsymbol{k}|\hat{\boldsymbol{S}}_{\mathbb{R}}|u^{\mathbb{R}}_m, \boldsymbol{k}\rangle |u_n^{\mathbb{R}}, \boldsymbol{k}\rangle \langle u^{\mathbb{R}}_m, \boldsymbol{k}| = V_{\mathbb{R}}\hat{\boldsymbol{S}}_PV_{\mathbb{R}}^{\dagger}
\end{equation}
With:
\begin{equation}
    \hat{\boldsymbol{S}}_{\mathbb{R}} = V_{\mathbb{R}}\hat{\boldsymbol{S}}V_{\mathbb{R}}^{\dagger}
\end{equation}
Being the spin operator in the real basis. Specializing to the z-direction, this spin operator enjoys particle-hole symmetry:
\begin{equation}
    \hat{S}_{P,z,\mathbb{R}} = -\hat{S}^*_{P,z,\mathbb{R}}
\end{equation}
By virtue of $\hat{U}_{C_{2z}\mathcal{T}, \mathbb{R}}$ acting diagonally. 
We can similarly define the reduced spin operator in this basis:
\begin{equation}
    \hat{S}_{P,z,\mathbb{R}}^r = \mathcal{U}_{ \mathbb{R}}^{\dagger} \hat{S}_{P,z,\mathbb{R}} \mathcal{U}_{ \mathbb{R}}
\end{equation}
Where $\mathcal{U}_{\mathbb{R}}(\boldsymbol{k}) = (|u^{\mathbb{R}}_1, \boldsymbol{k}\rangle, |u^{\mathbb{R}}_2, \boldsymbol{k}\rangle, \dots |u^{\mathbb{R}}_{N_{\mathrm{occ}}}, \boldsymbol{k}\rangle)$. This operator will also satisfy $\hat{S}_{P,z,\mathbb{R}}^r = -\hat{S}_{P,z,\mathbb{R}}^{r*}$, due to the diagonal action of $\hat{U}_{C_{2z}\mathcal{T}, \mathbb{R}}$ and the reality of the eigenstates because:
\begin{equation}
    \breve{W}_{C_{2z}\mathcal{T}, \mathbb{R}} = \mathcal{U}_{ \mathbb{R}}^{\dagger} \hat{U}_{C_{2z}\mathcal{T}, \mathbb{R}}\mathcal{U}_{\mathbb{R}}^{*} = \mathcal{U}_{\mathbb{R}}^{T} \mathcal{U}_{\mathbb{R}} = \mathbb{1}
\end{equation}
Thus, $\hat{S}_{P,z,\mathbb{R}}^r$ is an antisymmetric matrix related to the reduced spin operator in the original band basis by:
\begin{equation}
    [\hat{S}^r_{P,z,\mathbb{R}}]_{nm} = e^{-i(\varphi_n-\varphi_m)} [\hat{S}_{P,z}^r ]_{nm}
\end{equation}
Where $\varphi_{n}$ are phases such that:
\begin{equation}
    |u_n^{\mathbb{R}}, \boldsymbol{k}\rangle = e^{i\varphi_n(\boldsymbol{k})}V_{\mathbb{R}}|u_n, \boldsymbol{k}\rangle
\end{equation}
This is again a local (not global) unitary equivalence, but in a smooth gauge we expect them to not add additional topological structure. Thus, by choosing a smooth gauge $|\tilde{u}_1^{\mathbb{R}}\rangle$, we can get a well-defined Pfaffian invariant.  

We can fix the gauge at $\Gamma$ by recognizing that $H$ and $\hat{S}_z$ commute at $\Gamma$ in our two-band subspace, so that we can choose simultaneous eigenstates of $H$ and $\hat{S}_z$, which defines the rotation-symmetric state. Similarly, we can choose simultaneous eigenstates of $H_{\mathbb{R}}$ and $\hat{S}_{z, \mathbb{R}}$, which will be real as both matrices are real. Starting from this basis, we can then perform a local band mixing by taking the SVD of the overlap matrices, to get a gauge transformation $U$. This will maintain the chiral symmmetry, as long as $U$ is purely real, which it will be if we start from real eigenstates. This follows from gauge covariance:
\begin{equation}
    U^{\dagger}S_{P,z,\mathbb{R}}^rU = -U^{\dagger}S^{r*}_{P,z, \mathbb{R}}U = -(U^{\dagger}S_{P,z, \mathbb{R}}^rU)^*
\end{equation}
Note that this does not quite correspond to choosing a smooth gauge, as we have not ironed out the discontinuity at the end of the loop, arising from the non-trivial Berry phase \cite{Vanderbilt_smooth_gauge}. Such an ironing out procedure would break the reality condition. Luckily, this is not necessary for the Wilson loop quantization argument, which only relies on the Wilson loop being diagonal.

In this basis, finally, the Pfaffian will be a smooth function of $\boldsymbol{k}$ which changes sing across a nodal line. When $N_{\mathrm{occ}} = 2$ the Pfaffian is exactly equal to the spin eigenvalue of one band, tracked smoothly through any degeneracies. However, the Pfaffian also explains the topological protection of the nodal line more generally.

We note finally, that we find numerically that there exists  an alternative choice of gauge, without going through the real basis, where $\hat{S}_{P,z}^r$ has a global particle-hole symmetry, at least in the tetragonal magnetic case. At $\Gamma$, we can choose states such that the rotation operator is diagonal and so that $\varphi_1 = \varphi_2$. In this case, $C_{2z}\mathcal{T}$ is represented by $-\sigma_x$ at $\Gamma$. Parallel-transporting these states, we find that $C_{2z}\mathcal{T}$ is given by $-\sigma_x$ along the entire path $\Gamma$ to M.

\section{Further information on the models}\label{ap:model_details}
\subsection{Hexagonal system}
The models for the sixfold non-magnetic case were first considered in \cite{bouhon2019wilson}. They were constructed using a generalization of the Dresselhaus method to construct tight-binding models for a specific crystallographic space group \cite{dresselhaus2007group}. Specifically, they are constructed by starting with layer-group 77 (corresponding to the $z = 0$ plane of P6mm$1$'), and placing a $p_z$-orbital with both spin-components at Wyckoff position 2b, and then expanding the Hamiltonian, including all symmetry-allowed terms. This gives four bands, with twofold Kramers degeneracies at the TRIMs $\Gamma$ and M. We focus on gapped phases, where there are two occupied and two unoccupied bands.

When we only consider nearest-neighbor hoppings, we find a phase whose IRREP content is compatible with a fragile phase \cite{Ft1,bouhon2019wilson}. This is confirmed numerically in \cite{bouhon2019wilson} and Fig.~\ref{fig:extended_WL} by showing that the Wilson loop of both the occupied and the unoccupied subspace winds, but that they can be unwound when embedded in a higher-dimensional space. By contrast, when going up to 10th order neighbors in the hoppings, Ref.\cite{bouhon2019wilson} finds a phase that is still gapped over the BZ but now the IRREPs at $\Gamma$ have inverted. This phase onyl displays winding in the upper subspace. We note that in either case, a simple IRREP analysis would suggest that one of the band subspaces corresponds to an obstructed atomic limit, whereas the other one corresponds to a fragile limit, though whether the occupied or unocupied subspce is fragile/atomic obstructed is interchanged between the two phases. These phases are depicted in Fig.~\ref{fig:C6_combined}.

In the phase with few hoppings, Ref.~\cite{bouhon2019wilson} finds that the Wilson loop winds by $2$ over the BZ in both subspaces. By contrast, in the phase with long-range hoppings, the Wilson loop does not wind in the lower subspace, but winds by $4$ in the upper subspace. This difference is not captured in an IRREP analysis, but can be detected from the spin texture. In particular we find that the spin texture is trivial for both subspaces in the phase with Wilson loop windings of $2$, whereas it is nodal for both subspaces in the phase with Wilson loop windings of $0$ and $4$. Mote that the Wilson loop winding of $4$ for the upper subspace seems to contradict the results in Tab.~\ref{tab:pseudo_spin_Wilson}. However, those results assume that there is a twofold degeneracy at $\Gamma$ and K, which is only satisfied for the lower subspaces in Fig.~\ref{fig:C6_combined}. Thus, the upper subspace has a nodal spin texture, but this is not captured by our link to connectivities between degeneracies. 

\subsection{Tetragonal system}
The tetragonal system with winding Wilson loop was first introduced in Ref.~\cite{Bouhon2021}. Starting from magnetic space group P$_{\mathrm{C}}$4, (No. 75.5 in the BNS convention), we place an s-orbital with both spin-components at Wyckoff position 2b, again giving $4$ bands. There are twofold Kramers degeneracies at $\Gamma$ and M, but not at X due to the magnetic structure of the space group. Thus, the lowest possible connectivity is $2$. We again construct a model using a generalized Dresselhaus procedure, though here we also use \texttt{MagneticTB} \cite{MagneticTB} for easier construction of higher-order hopping models.

We again focus on the case where there are two occupied and two unoccupied bands that are gapped across the BZ (for a discussion of other phases, see \cite{Bouhon2021,Lange2021}). For nearest-neighbor hoppings, we find a phase with Wilson loop winding $1$ both subspaces. Going up to the tenth order hoppings, we find a phase where the IRREPs at X have inverted and the occupied subspace has Wilson loop winding $0$, whereas the unoccupied subspace has Wilson loop winding $2$. Similar to the hexagonal case, a pure IRREP analysis would suggest that in either case there is one fragile subspace and one atomic obstructed subspace, though they again interchange.

In the case where the Wilson loop winding is $1$ in either subspace, the spin texture is trivial across the BZ. However, when the windings are $0$ in the occupied and $2$ in the unoccupied, we find a nodal-line spin structure in both subspaces, as shown in Fig.~\ref{fig:C4_combined} (though see the comment about the Wilson loop winding of $0$ in Appendix~\ref{ap:C4_results}).

\section{Results for the fourfold magnetic case}\label{ap:C4_results}
\begin{figure*}[ht!]
    \centering
    \includegraphics[width=\textwidth]{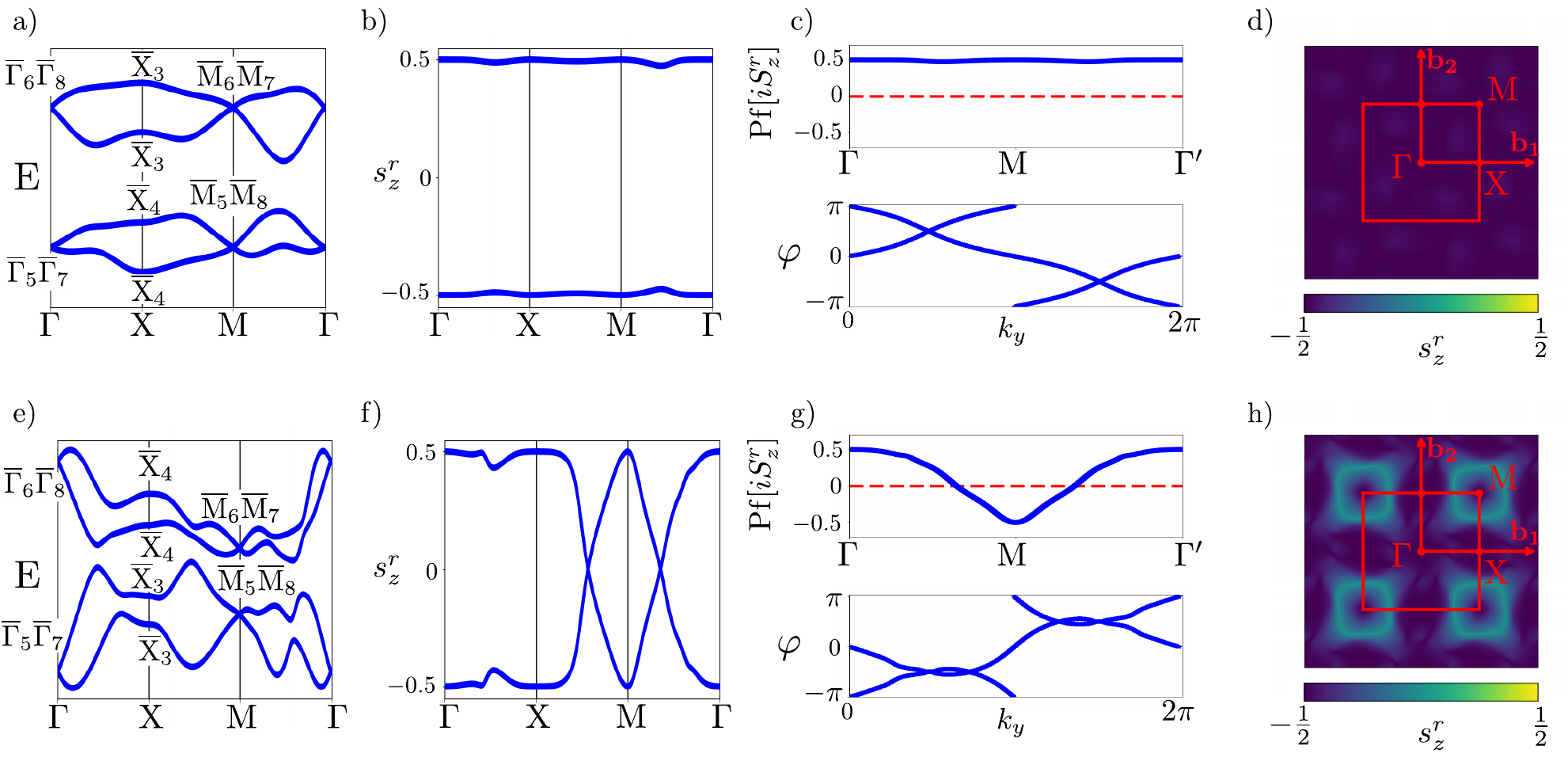}
    \caption{Relevant figures for the P$_{\mathrm{C}}4$ symmetric case (similar to Fig.~\ref{fig:C6_combined}). In a)-d) we show the fragile case (where the Wilson loop of both subspaces winds), whereas e)-h) shows the trivial case (where only the occupied subspace has a winding Wilson loop). In a) and e) we show the band structure, with IRREPs of P$_{\mathrm{C}}4$ indicated. In b) and f) we show the eigenvalues of the reduced spin operator along high-symmetry paths of the BZ. Note that spin is not a good quantum number at X, but is a good quantum number at $\Gamma$ and M. In c) and g), we show the Pfaffian and Wilson loop over the BZ of the occupied bands, where $\Gamma' = \Gamma+\boldsymbol{b}_1$. The Wilson loop winds in c) but not in g) (see comment in Appendix~\ref{ap:C4_results}). Finally in d) and h) we show the spin expectation value over the BZ of the lowest spin band [as shown along high-symmetry lines in b) and f)]. This illustrates that when the Wilson loop winding is even (including $0$) the system displays spin nodal lines. For more details on the models, see Appendix.~\ref{ap:model_details} and Ref.~\cite{Bouhon2021}.}
    \label{fig:C4_combined}
\end{figure*}

We show the results for the fourfold magnetic case in Fig.~\ref{fig:C4_combined}, in complete analog to Fig.~\ref{fig:C6_combined}. Note that it is difficult to tell the winding of the lower subspace [Fig.~\ref{fig:C4_combined}g)], as it is not clear whether the Wilson bands form an avoided crossing. We can check this by choosing a different Wilson loop geometry (note that whether or not the Wilson loop winds should be independent of geometry). We therefore look at the winding as we deform our base path in a $C_4$ symmetric manner, starting from the path $\Gamma$M $\Gamma'$ with $\Gamma'=\Gamma+\boldsymbol{b}_1$ and smoothly deforming it to the path $\Gamma$X$\Gamma'$. This covers a quarter of the BZ, and should therefore correspond to 1/4 of the total winding (we considered the same construction in \cite{Bouhon2021}). In doing so, we find the expected result: The Wilson loop does not wind in Fig.~\ref{fig:C4_combined}g), so that the Wilson bands form an avoided crossing. 

\section{Spin Wilson loops}\label{ap:Spin_Wilson_loops}
\begin{figure*}[ht!]
    \centering
    \includegraphics[width=\textwidth]{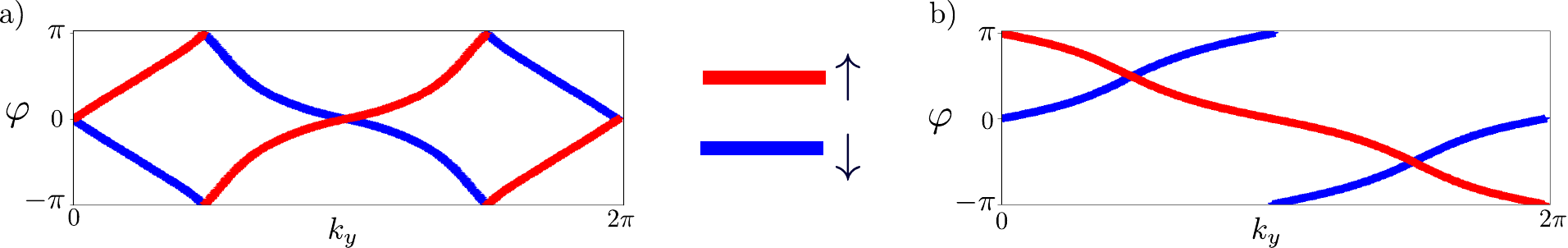}
    \caption{Wilson loop winding for occupied band-subspace projected onto spin-space, with colors indicating the component of spin responsible for the winding. a) Shows the hexagonal case, whereas b) shows the tetragonal case. In a), we have a spin Chern number of 4, whereas in b) we have a spin-Chern number of 2} 
    \label{fig:spin_wilson_loop_winding}
\end{figure*}

\begin{figure*}[ht!]
    \centering
    \includegraphics[width=\textwidth]{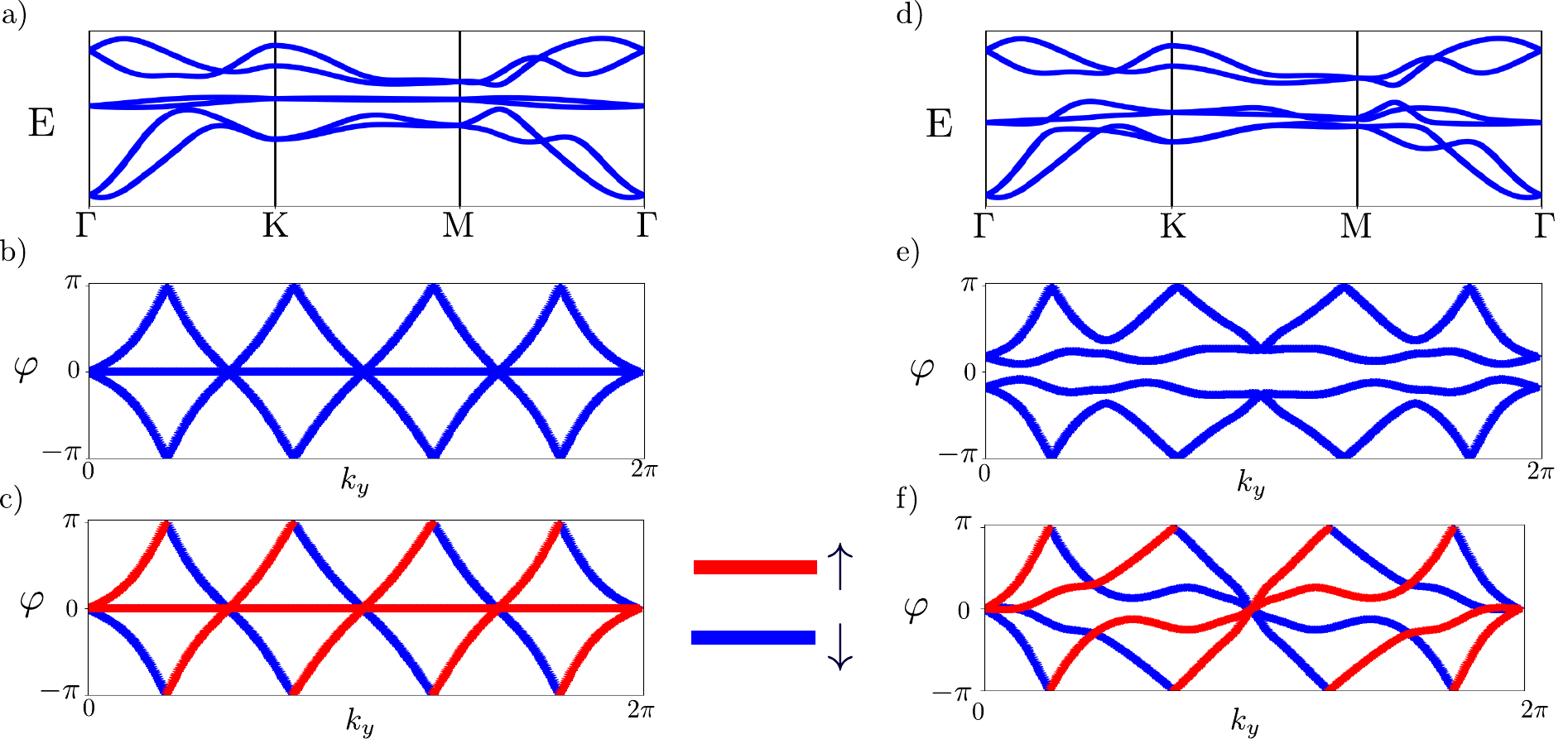}
    \caption{Band topology of hexagonal system when coupled to two trivial DOFs (for details of this coupling, see Ref.~\cite{bouhon2019wilson}). This is for the case where, originally, both the occupied and the unoccupied space have a winding of $4$. We consider the topology of the four top bands. We check that the energy gap to the lower two bands, and the spin gap of the four-dimensional band subspace, is open across the BZ. In a)-c) we show a case where the trivial bands are not coupled to the winding bands. We show the band structure in a), with bands with trivial winding in the middle. In b), we show the Wilson loop spectrum for the top four bands, which is decoupled in this limit. In c), we show the spin Wilson loop, for the two top/bottom spin bands in red/blue respectively, with a total spin-Chern number of 8. In d)-f), we show the same results when coupling the bands with trivial and non-trivial winding. We see that the Wilson loop gap, as expected in fragile topology. However, the spin-Wilson loop does not gap, suggesting that the spin-Chern number is preserved. Note, however, that it maps to a trivial Kane-Mele invariant in either case \cite{KaneMeleZ2}.} 
    \label{fig:extended_WL}
\end{figure*}

Whenever the spin spectrum is gapped (as happens in our models when the pseudo-spin connectivities are trivial and the electronic band topology is non-trivial), we can compute spin-resolved Wilson loops. Following \cite{Lin2022}, we compute the Wilson loops projected into the upper and lower spin bands respectively. We present the results in Fig.~\ref{fig:spin_wilson_loop_winding} for the lower electronic subspace only, though we find the same results for the upper electronic subspace. For completeness, we also show the spin Wilson loops for a case where we directly unwind the Wilson loop in Fig.~\ref{fig:extended_WL}.

\end{document}